\documentclass{article}
\usepackage[utf8]{inputenc}
\usepackage{multicol}
\usepackage{blindtext}
\usepackage[margin=1in]{geometry}
\usepackage{titlesec}
\usepackage{graphicx}
\usepackage[english]{babel}

\title{Tired of Misattribution, Modeling Player Fatigue in the NBA}
\author{Austin Stephen, Matthew Yep, Grace Fain, Ron Yurko, Maksim Horowitz}
\date{November 2021}
\titleformat*{\section}{\large\bfseries}
\titleformat*{\subsection}{\normalsize\bfseries}
\titlespacing*{\subsection}
{0pt}{3ex plus 1ex minus .2ex}{3ex plus .2ex}
\begin{document}

\begin{titlepage}
   \begin{center}
        \vspace*{1cm}
        
        \LARGE
        \textbf{Tired of Misattribution, Modeling Player Fatigue in the NBA}
        
        \vspace{4cm}
        \large
        \textbf{Austin Stephen$^1$, Matthew Yep$^2$, Grace Fain$^3$}
        \vspace{.1cm}
        
        Advised by Ron Yurko$^4$, Maksim Horowitz$^5$
        \vspace{.75cm}
        
        \normalsize
        University of Wyoming$^1$, University of California Berkeley$^2$,University of Oklahoma$^3$
        Carnegie Mellon University$^4$, Atlanta Hawks$^5$

    \end{center}
\end{titlepage}

\begin{multicols*}{2}
\begin{flushleft}
\section{Abstract}
\setlength{\parindent}{10ex}The prevailing belief propagated by NBA league observers is that the workload of the NBA season dramatically influences a player's performance$^{1,2,3,4,5}$. We offer an analysis of cross game player fatigue that calls into question the empirical validity of these claims. The analysis is split into three observational studies on prior NBA seasons. First, to offer an analysis generalized to the whole league, we conduct an examination of relative workloads with in-game player tracking data as a proxy for exertion. Second, to introduce a more granular perspective, we conduct a case study of the effectiveness of load management for Kawhi Leonard. Third, to extend the analysis to a broader set of fatigue sources, we examine the impact of schedule features structurally imposed on teams. All three analyses indicate the impact of cumulative player fatigue on game outcomes is minimal. As a result, we assert in the context of the measures already taken by teams to maintain the health of their players,$^6$ the impact of fatigue in the NBA is overstated.

\section{Status Quo NBA} 
Throughout the past decade, addressing player fatigue at a schedule level has been and continues to be a prevalent concern for both NBA teams and the league office.

\subsection{Density Reduction}
One way the league office has reduced the burden placed on athletes is by removing occurrences where teams play four games in five days. In the 2012-13 season, there were 76 instances of four in five, in 2016-17 such instances were reduced to 23, and by the 2017-18 season, four games in five days became extinct from the schedule. This was achieved by strategically starting the season a week earlier, such that increasing the season length from 170 to 177 days allowed for a less condensed schedule$^7$. Such efforts have also aimed to reduce the number of three games in four days as well as back-to-back games, however these events are still present throughout the schedule. 

As of the 2020-21 NBA season, the league office also introduced a new feature to the schedule: two-game series, where visiting teams play the home team two times in a row at the same arena. These series are played over the course of either two or three days. Not only did they reduce traveling and impose restrictions to reduce COVID-19 exposure, they also decreased the amount of one game road trips in the schedule. 

However, despite the league’s efforts to design less densely packed schedules, many coaches such as Greg Popovich and Nick Nurse have still adopted load management strategies where they rest their key players during the regular season. This is problematic to the NBA league office, especially when these star-less games are nationally televised. Not only are these games less entertaining for fans, they consequently result in substantially lower television ratings$^8$.

To address this, in recent years the league office has instituted a resting policy that prohibits teams from resting healthy players in high profile, nationally televised games. In addition, barring unusual circumstances, teams should not rest multiple healthy players in the same game, nor should they rest healthy players on the road. The fines for violating these resting policies sit at a minimum of \$100,000$^9$.

\subsection{Schedule Fairness}
Many NBA teams located on the west coast and southern borders travel substantially more miles throughout a given season compared to northeastern teams. Consequently, these teams also cross many more time zones to get to and from games. \\

\begin{center}
\includegraphics[width=6cm]{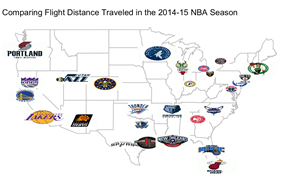}\newline
Fig. 1
\end{center}

The map above illustrates the cumulative flight distance traveled for each NBA team in the 2014-2015 season. The sizes of the logos are scaled using percentiles, showing that the larger the logo, the more that team had to travel. The Portland Trail Blazers had to fly over 62,000 miles that season, while the Cleveland Cavaliers only traveled 35,000 miles. 

This raises an important question, to both NBA teams and the league office, as to whether or not the teams who travel more face a structural disadvantage due to scheduling. Aggregated travel could inflict cumulative fatigue, and having to cross time zones could also very well disrupt players’ circadian rhythms and inhibit their performance and ability to recover. 

\section{Related Work}
\subsection{Defining fatigue in basketball}
To maintain consistency with prior work examining fatigue in basketball$^{10}$, we define fatigue as “an acute impairment of performance that includes both an increase in the perceived effort necessary to exert a desired force and the eventual inability to produce this force”$^{11}$. This definition is inclusive of fatigue from any source not discriminating against muscular cognitive or any other qualifier attached to the source of the fatigue. On average, during a game of basketball, players travel 5–6 km at physiological intensities above lactate threshold and 85\% of maximal heart rate$^{12}$. This leaves no question basketball competitions levy a degree of strain on players that given insufficient recovery could induce fatigue under this definition.
\subsection{Fatigue’s and athlete performance}
There is an extensive body of work examining the physiological demands placed on athletes in basketball at all levels. A literature review of twenty-five publications found evidence that players incur fatigue as the game progresses, albeit to differing degrees depending on position$^{12}$. “Higher-level players”, the category NBA players would fall into, sustain greater workloads than lower-level players at the same positions$^{12}$. This suggests the NBA is a prime candidate for examining fatigue accumulation over a season. 

Nonetheless, the literature also creates a basis for questioning prevailing beliefs on fatigue’s influence on game outcomes because several research groups have failed to find a relationship between the muscular properties of athletes and player performance in basketball. An analysis of eight semi-professional basketball players found “trivial-small associations between all external and internal workload variables and player efficiency”$^{13}$. Furthermore, an article from Barca Innovation Hub examining training workload found no connection between a higher external load and better performance in-game$^{14}$.

As opposed to in-game fatigue, analyses on structural fatigue sources suggest fatigue does influence performance. A research group using t-tests and analysis of variance (ANOVA) found a significant difference of means in traveling West to East asserting this influences a player’s circadian rhythm$^{15}$. It is well-established that air travel in general influences athletes and one research group asserts these impacts should materialize in the NBA$^{16}$.

One of the biggest performance concerns with athlete fatigue is potential for increased injuries with higher workloads. Work examining sports in general has shown that combining an understanding of the training dose and external influences helps players prevent injury and improve performance$^{17}$.  However, there are inconsistent results across the literature about the impact of fatigue on injury occurrence in basketball. A paper examining thirty-three NBA players in an observational retrospective cohort study found an inverse relationship between training workload and injuries$^{18}$. Yet, a work using over 70,000 NBA games and 1,600 injury events showed the likelihood of injury increased by 2.87\% for every ninety-six minutes played and decreased by 15.96\% for every day of rest$^{19}$. 
\subsection{Gap in the literature}
There are three main gaps in the literature targeted by the three sections of this paper. First, we are unaware of any research that measures relative player workloads and performance at a league wide level. This allows examination of exertion and cumulative fatigue related to in-game performance. These ideas are addressed in the section of the paper titled In-Game Workload Induced Fatigue. Second, we are unaware of any case studies on the effectiveness of strategies that rest players to avoid fatigue accumulation, referred to as load management colloquially. This is explored in the section titled Case study of Kawhi Leonard. Lastly, while there has been work on structural fatigue in basketball$^{15, 16}$, we propose a more robust procedure for isolating the impact of structural fatigue that is less vulnerable to latent variables and offers interpretations on the relevance of observed impacts to game outcome. This work is detailed in the section titled Structural Fatigue.

\section{Data}
To access data about player and team performance, as well as potential sources of fatigue, we used a set of existing R packages maintained by the basketball analytics community. These packages include nbastatr$^{20}$, airball$^{21}$, and NBAr$^{22}$. They provided access to box score statistics about player performance and some more advanced metrics such as game-net rating and Player Impact Estimate at a game and season level. To measure in-game fatigue we used a player’s total in-game distance traveled via computer vision generated by Second Spectrum and courtesy of nbastatr$^{20}$. 
\section{In-game workload induced fatigue}
In-game workload induced fatigue examines player in-game workload and performance. The analysis of in-game workload induced fatigue consists of three additional conceptual divisions. First, an examination of how player exertion correlates with performance. Second, relating an accumulation of player workload over a window of time and player exertion in an observation game. Third, examining how accumulation of player workload over a window of time influences player performance. These are located in sections 5.3, 5.4 and 5.5 respectively.

\subsection{Theoretical foundations}
A player’s relative in-game distance is used as the proxy for player workload. The formal definition of work from kinematics includes an object's displacement and the force acting on the object. Where W is work, F is force, and s is displacement.
$$W = Fs$$
The force to move an object of constant mass is also constant. Assuming a player’s mass is constant over a season, changes in distance for a game captures the relative change in that player's work for the game. In other words, work is agnostic to the duration and cardinality of the player's motion and leveraging this allows measurement of player exertion with only tracking data of player displacement. It is important to acknowledge the literature establishes that in basketball different types of motions are associated with slightly different degrees of exertion due to the biomechanics associated with them$^{23, 24}$. Furthermore, players pushing against each other during games, moving in the z axis via jumps (of particular relevance to basketball), and the tracking data being generated by a computer vision model all introduce additional noise. However, there is no publicly available tracking data with the granularity to account for these sources of noise nor is there a guarantee they even cause enough noise to justify the complexity of designing a system of equivalence. Therefore, studying a player's total displacement in the xy plane relative to their own baseline offers the most effective picture of relative changes in work available.
\subsection{Exertion and performance methodology}
An important assumption baked into the hypothesis that a fatigued player’s performance suffers is that exerting more effort in a game positively affects performance. This relationship is non-trivial because the scoring dynamics of basketball are highly tied to athlete skill level offering some isolation from exertion not present in sports like track. 

The relationship between exertion and performance is modeled using simple linear regression as there was found no indication a more flexible method was necessary. In the models, a player’s deviation from their season average in a set of performance measures is the response and the player’s deviation from their season average in distance traveled divided by minutes played is the predictor. $$PMD =\beta_0 +\beta_1(DSD)$$

Where PMD is performance measure deviation from season mean,
DSD is distance per second of playtime deviation from season mean.
As discussed in the Theoretical foundations, a player’s deviations from their season average treats player mass as constant and correlates the relative change in work with a relative change in performance. The distance measured is the total player movement for the game (regardless if they have the ball). If the performance statistic measures offensive performance, such as field goal percentage, then only offensive distance divided by offensive minutes is used. This ideally tailors the players exertion in that subarea of the game to their performance in that subarea as well. It is important to note this procedure was not used when measuring total workload for fatigue discussed in the next section.
The distance divided by playtime is necessary because many cumulative statistics like rebounds and steals would naturally accumulate as a player is in the game for more time which is not necessarily indicative of better performance. 

Each observation in this analysis is a player in a game, and two subsets of this data were used to improve the quality of the findings. The first subset was composed of players who in the game four or more minutes, and the second was twelve or more minutes (NBA games are 48 minutes long). Excluding players in the game for fewer than four or twelve minutes reduces the variability in player performance measures. For example, a player who is substituted into the game for a couple minutes and misses their one shot  but in the next game makes their single shot would be a 100 percent change in shooting percentage. These small time periods offers a very noisy measurement of that player's performance. Filtering by minutes, as opposed to a different arbitrary cutoff, introduces less selection bias towards players with certain behavioral patterns. For example, a cutoff on number of shot attempts would introduce a bias towards players who take lots of shots. Two subsets are reported because a more strict limit on the number of minutes in the game lowers variability in performance measures at the tradeoff of only measuring players who are in the game for a larger percentage of the time. The results section highlights the importance of this procedure to the empirical
findings.
\subsection{Results exertion and performance}
There are two major takeaways from the analysis offered in this section with full results in Fig. 2 and Fig. 3. The first takeaway is deviations in player exertion appear to be loosely positively correlated with their performance in the game. Points scored is arguably the most important performance metric because the points a player scores is most closely tied to the game’s outcome. Points scored shows a strong correlation with player exertion for both subsets, p-values of .01 and 1e-12 for the four and twelve minute subsets respectively. This offers evidence of a positive relationship between exertion and scoring. Furthermore, the correlation was positive and statistically significant for field goal percentage, offensive rebounds, and steals across both subsets.\newline 
\vspace{4cm}
\begin{center}
\includegraphics[width=6cm]{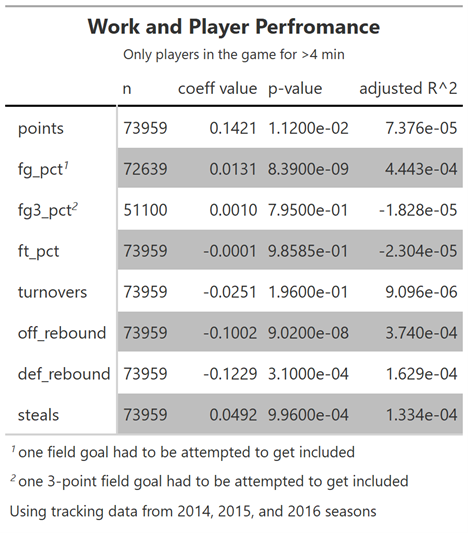}\newline
Fig. 2
\end{center}
\vspace{.5cm}
\begin{center}
\includegraphics[width=6cm]{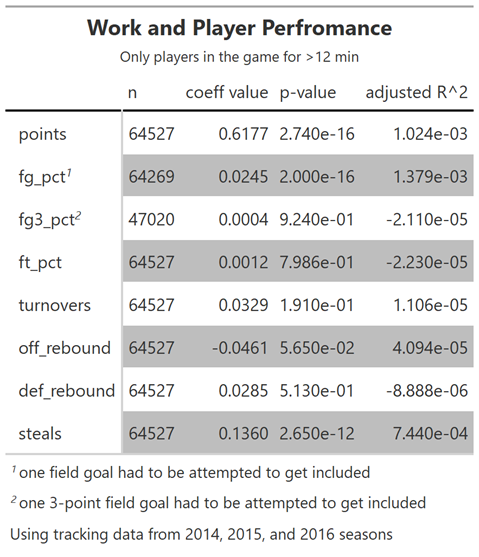} \newline
Fig. 3
\end{center}

Despite these relationships, the results present a reasonable degree of ambiguity about how exertion relates to more granular measures of performance. 
There was no correlation in at least one subset between turnovers, three point shooting, offensive rebounds, and free throws. 
Furthermore, there was a significant inverse relationship with the number of offensive rebounds (more offensive rebounds is considered good in basketball) leaving a literal interpretation that greater exertion hurts performance in this capacity. 
Furthermore, the response is highly sensitive to the subset of the data used.
P-values varied by several orders of magnitude across the subsets. 
Despite achieving significant p-values on some measures the dramatic change in results across subsets has two potential causes. 
First, it could suggests their is a different underlying relationship between the these different areas of the data or second, the large number of observations is leading models to over assert the prominence of weak relationships, an established problem with large sample sizes and the p-value $^{25}$. 
Either interpretation raises questions about how any conclusions regarding exertion and performance will generalise.

The second major takeaway is exertion accounts for a low percentage of variation in performance across all measures and time constraints. Players in the game for longer than 12 minutes and field goal percentage had the largest adjusted-r squared value (.0013) of all measured factors. However, even this accounted for less than 1 percent of the variation in the response. This has two potential interpretations. First, player exertion accounts for a very small, arguably negligible, amount of the variation in player performance. Alternatively second, measuring player exertion in this fashion introduces too much noise to capture the magnitude of its relationship with performance. Without additional experimentation it is challenging to argue for either hypothesis beyond speculation, however, there is work in the literature that has shown there is some ambiguity attached to the notion simply exerting more work means an athlete will perform better in basketball$^{14,18}$.
\subsection{Post distance overload regress}
If players experience a performance detriment associated with fatigue, then subsets of the season where a player’s workload increases should be followed by a regression. Extracting this relationship required examination of windows of the NBA schedule at the player level. An observation included a player in the game for more than 15 minutes for the 2015 season (n= 19,751). The playtime cutoff of 15 minutes intends to exclude players who are in the game for a short period of time as it is unlikely they experience measurable cross-game fatigue. To examine a potential association between prior workload and observed game performance a simple linear regression was used. The previous x days difference from the players season average in distance was treated as the predictor. The distance in the game difference from season average immediately following the window of time was treated as the response. Fig. 4 contains a scatter plot of these results for window sizes of 3, 5, 7, and 10 days with the regression line.

\begin{center}
\includegraphics[width=6cm]{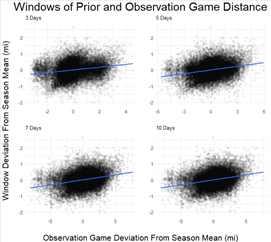} \newline
Fig. 4
\end{center}

This analysis offers evidence contrary to what the fatigued player hypothesis would suggest. Players that covered more distance in prior games covered more distance in the observation game as well across all windows. These relationships had $\le$2e-16 p-values and adjusted r-squared values showing about 5 percent of the variance in the response is explained by the window. One hypothesis for such a strong correlation is this measurement strategy ties too closely to player’s usage rate which maintains local temporal variations as their role on the team evolves over the season. This can be adjusted for by using a playtime adjusted response which is agnostic to a players usage rate. The trade-off of this analysis is measuring distance playtime adjusted is no longer a proxy for player work. Scatter plots of the relationships for windows sizes 5 and 7 are in Fig.6 , and show how the positive association disappears, supporting this hypothesis. \newline 

\begin{center}
\includegraphics[width=6cm]{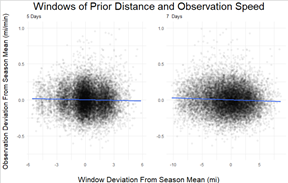} \newline
Fig. 5
\end{center}
\vspace{.25cm}
In fact, the $\beta_1$ coefficient for the linear regressions became negative with p-values of .003 and 1.1 e-6 for windows of five and seven days respectively. This indicates an albeit small significant negative association. A concern with such a small the negative relationship is it could simply be regression to the mean. Furthermore, making the playtime adjustment violates the assumptions set in the empirical motivations as it no longer measures cumulative fatigue. This makes the interpretation of this result in the context of cumulative fatigue impossible. 
\subsection{Post distance overload performance}
Lastly, searching for any highly compelling evidence of players experiencing a workload regression, we ignored all intermediate measures and directly compared prior workload to observations of game outcomes. Fig. 6  and 7 shows the results of treating performance measures as the response and windows remain the predictor. 

In line with the prior results the relationship between these performance measures and fatigue is ambiguous. Field goal percentage achieves a significant p-value in the 3 day window but no others shown in Fig. 6. Furthermore, with points there is a positive association between prior distance, meaning a literal interpretation would suggest players who are more fatigued score more points. Similar to the first windowed analyses that found a positive association between prior workload and observation workload, the most likely hypothesis for the relationship is the usage rate increases as players are performing better. Therefore, if a player is scoring more points in prior games they are given more playtime and in turn travel more distance.
\begin{center}
\includegraphics[width=6cm]{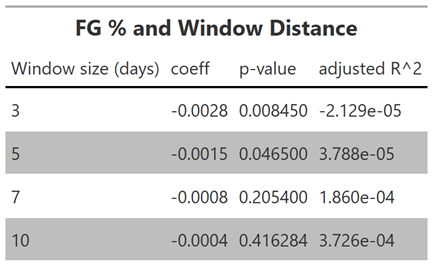} \newline
Fig. 6
\end{center}
\vspace{.25cm}
\begin{center}
\includegraphics[width=6cm]{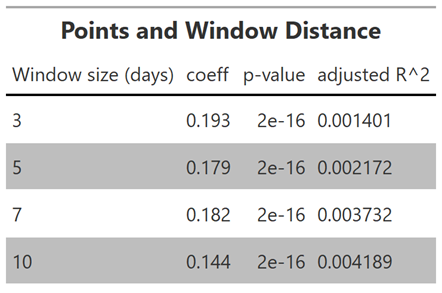} \newline
Fig. 7
\end{center}
\subsection{In-game workload big picture}
Across all of the approaches to modeling fatigue influences on performance there is a consistent narrative. Regardless how the data is subset,  normalized, or what is treated as the response there appears to be an inconsistent and largely ambiguous relationship between fatigue and performance.
\section{Case Study of Kawhi Leonard}
Load management is a strategy employed by NBA coaches where they deliberately rest players, often star players, from playing a game, with the goals of reducing their risk for injury and preserving them for the playoffs. 
While these are ultimately long run goals, the case study below explores how strategies of load management can pay off in the short run. In particular, it delves into whether or not Kawhi Leonard, the NBA player most notorious for practicing load management, plays better in the games immediately following periods of extra rest. 
\vspace{.25cm}
\begin{center}
\includegraphics[width=6cm]{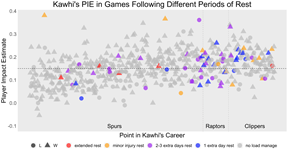} \newline
Fig. 8
\end{center}
\vspace{.25cm}
Each point in Fig. 8 represents a game in Leonard’s career, with the blue points showing instances where he was load managed before the game versus the gray points showing no extra days of rest. The response variable is Player Impact Estimate; PIE is the NBA’s metric for player performance, factoring in both defense and offense. The dashed line across the middle represents an exceptional PIE of 0.15, which was the 95th percentile in the most recent NBA season.

Using a difference of means test, Kawhi was no more likely to perform better after being load managed. This is visually apparent in Fig. 8.
The majority of all of the blue points are above the dashed line, suggesting that Leonard plays exceptionally well after resting for an extra day. However, these blue points are not convincingly above the gray points, as many of the gray points are also located above this dashed line. Even in games following no extra rest, Leonard performs at an elite level for his team. Thus there exists no substantial evidence to support that Leonard necessarily performs better in the games following load management versus the games where he did not rest extra. This suggests that load management is not particularly effective in the short run, at least for Kawhi Leonard. 
\section{Structural Fatigue}
Structural fatigue encapsulates any factors outside of a team’s control, in other words, induced by the structure of the schedule. For example, teams have no influence over how far they have to travel to play a game or how many games they play in a week. We aggregated travel data from the 2010 through 2019 NBA seasons, courtesy of airball$^{21}$. This data consisted of between-game potential features of interest such as time zone changes, tipoff time, flight distance, days of rest and schedule density. 
 This section is an examination of how these schedule-induced-factors influence performance at a team level.

\subsection{Model Construction}
The first step to developing the model was to find a strong measure for capturing game outcome. Compared to alternatives such as score differential or a binary measure like win loss, game net rating offered the most detailed and accurate measurement of the result of a game. Game net rating is the difference between a team’s offensive and defensive rating for a given game, and both of these ratings are calculated per 100 possessions. Having the response variable standardized to be per 100 possessions allowed for stronger comparisons across NBA seasons as playstyles and points per game evolve over time.
 
Measuring the effects that schedule induced factors have on game net ratings required normalizing for the strength of the teams. 
There was significant exploration for the strongest proxy of team strength. The simplest model uses the difference of the teams’ end of season net rating. Attempting to improve on this, models that temporally weight games based on recency, nonparametric models including random forest, and dimension reduction techniques over a large set of team information using elastic net were all tested. These models all achieved improvements within 2 standard errors in 10 fold cross validation of team net rating. As this paper was not about modeling team strength, an active area of sports research, we selected end of season net rating in a linear regression model as a proxy for team strength. We acknowledge improving the proxy for team strength is an excellent direction to expand the results of this work.

After determining a proxy to account for differences in team strength, the next step was developing a model in two stages. The first stage regressed game net rating on the net rating difference strength proxy. Then by calculating the residuals of this first regression and regressing them on all of the travel metrics, the second stage effectively teased out what portion of the variance in game net rating the travel metrics could explain that the strength proxy could not. The results can be seen in the following figures. 
\subsection{First model results}
$$game\: net\: rating = \beta_0 + \beta_1(net \:rating\: di\!f\!ference)$$
The proxy for team strength difference, net rating difference, yielded an adjusted R-squared of 0.2262. The data was structured such that the response was relative to the visiting team, meaning the $\beta_0$ term captures that on average, visiting teams are disadvantaged by about -2.81 in game net rating. The 1coefficient presents that for each point in net rating that a given team had more than their opponent, their game net rating for that game is expected to increase by 0.968. 

\subsection{Second model results}
$$ residuals = \beta_0 + \beta_i (schedule factor_i)$$

\begin{center}
\includegraphics[width=6cm]{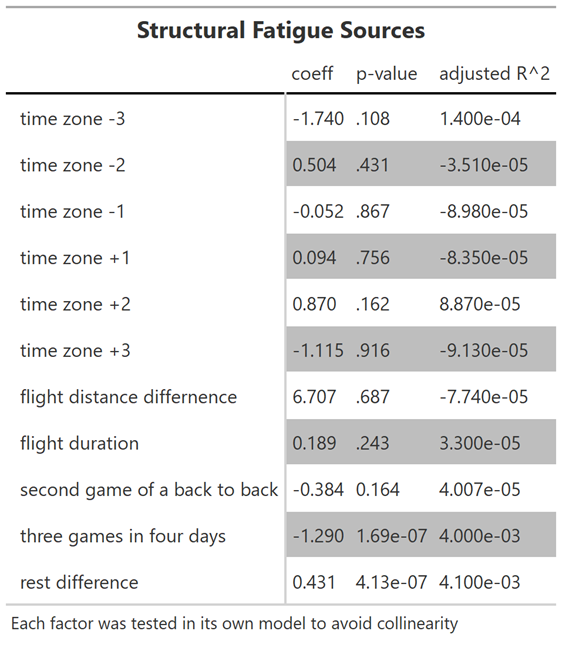} \newline
Fig. 10
\end{center}

The regression results convey that out of all of the travel metrics, days of rest difference, traveling three hours back, and the observed game being the third game in four days were the only variables that had statistically significant effects on game net rating. The coefficient on rest difference presents that the visiting team’s game net rating is expected to increase by 0.35 for each day of rest they have more than their opponent. Traveling three hours back also had a detrimental effect; if a team had to travel from the east coast all the way west, their game net rating is expected to go down by about -1.740. Likewise, if the game being played is the team’s third game in four days, the model estimates their game net rating to decrease by -1.290. However, it is important to acknowledge these results had adjusted R-squared indicating they explain less than .1 percent of the variance in the response. Meaning they offer little information about why a game turned out the way it did.

It was perhaps more insightful, however, to observe that many of the collected travel metrics had insignificant effects on game net rating. Contrary to conventional wisdom, the observed game being the second leg of a back to back did not have a significant negative effect. Neither did the majority of our time zone shift metrics, save for three hours back. In addition, all of the distance metrics- team flight distance, flight distance difference, and windowed versions of these metrics were also insignificant. This addresses the aforementioned concern that having to travel and cross more time zones inherently puts teams at a disadvantage. The results offer that this is not the case. Moreover, this provides incredibly useful insight to the NBA league office because it suggests that scheduling teams to fly more miles does not necessarily put them at a structural disadvantage, but that scheduling them to have 3 games in 4 days appears to be more consequential to performance. Such insight can help the league office effectively design schedules that are more fair. 
\section{Relevance to Stakeholders}
Fatigue analysis provides valuable insight to NBA teams and the league office, from strategic, schedule fairness, as well as viewer and profit maximization standpoints. 

\subsection{NBA teams}
NBA teams can utilize fatigue analysis to understand how to best allocate their resources and important areas of focus for the team. When teams, trainers, and athletes understand how fatigue is impacting each game, they can develop strategies to minimize fatigue. A team can use this knowledge to reallocate time and effort, such as changing team travel plans, changing how and when the team gets to the stadium, or adjusting player’s rest to practice/warm-up ratio in order to decrease fatigue's impact on game net rating. Our work in particular suggests that teams' existing strategies are highly effective at mitigating the impacts of fatigue. Across all analyses the impact of player fatigue on game outcomes was ambiguous with the exception of playing 3 games in 4 days, and the rest difference between teams at a schedule level. These rare scheduling events may represent the last holdout of improvements that can be achieved by NBA teams. Therefore, from an optimizing player performance perspective there is little incentive for raised investments from teams into player recovery when existing procedures appear to be eliminating any regressions in performance.

The case study of Kawhi Leonard shows that the impact of load management was somewhat ambiguous, at least in the short run. An NBA team would want to perform a similar analysis with other key players. This is because there exists an opportunity cost every time a player does not suit up for the game. They will be more rested and have less potential for injury, but there is a higher chance of losing that game, and  other players will have to exert themselves more. Therefore it would be important for NBA teams to carry out some form of a cost-benefit analysis to explore whether or not the payoffs of load management outweigh the deficits. 
\subsection{NBA league office}
The structural fatigue analysis identified certain features of the schedule that detrimentally impacted team performance, albeit with a moderately small overall impact on game outcome. From a league office perspective, they want to strategically minimize occurrences of these events, but if they are unavoidable, the league will want to make sure that these events are distributed fairly so that certain teams are not put at structural disadvantages. 

Additionally, structural fatigue analysis is important when considering viewership and profit maximization. Using strategies like starting the season a week earlier as a blueprint, the league office can continue to take measures that adjust the schedule so that players have adequate time to rest and recover between games. Schedule analysis drives insight to inform what kind of changes need to be made in order to do so. This kills two birds with one stone as it also serves as a means of discouraging teams from load managing their players. Organizing the schedule in such a way that coaches like Popovich or Nurse do not even have the need to rest their stars could effectively ensure that marquee players are suited up on the court. Doing so would very well increase the entertainment value of NBA games, which in turn boosts broadcasting viewership and revenue generated. 
\section{Future Work}
\subsection{Exertion and performance}
In the section Results on Exertion and Athlete Performance, we are unable to establish a string association between player exertion and performance. There is almost certainly a more pronounced relationship between player performance and exertion. In particular, there is significant room to improve this analysis with more granular tracking data that better characterizes in-game events. For example, off the ball movement requires less energy than on ball movement and more accurately reflects the game. Achieving better mappings between the muscular properties of athletes and game outcomes will create a more nuanced picture of how exertion influences performance. At a team level, this picture could be improved by incorporating practice workload into the picture to build the most accurate measure of cumulative workload.
\subsection{Long run load management}
The case study suggests that Kawhi Leonard’s player impact estimate does not necessarily improve in games immediately following additional rest compared to those with no load management. This calls into question the effectiveness of load management as a strategy in the short run, however it would ultimately be more important to figure out how load management pays off in the long run. 
\section{References}
\begin{enumerate}
\item Holmes, B. “Schedule alert! Every game tired teams should lose this month.” ESPN Enterprises, Inc, 1 Nov. 2017, https://www.espn.com/nba/story/\_/id/\newline 21236405/nba-schedule-alert-20-games-tired-teams-lose-november.

\item Holmes, B. “Which games will your team lose because of the NBA schedule?.” ESPN Enterprises, Inc, 30 Oct. 2018, https://www.espn.com/nba/story/\_/id/\newline 25117649/nba-schedule-alert-games-your-team-lose-2018-19.

\item Magness, S. “Fatigue and the NBA Finals: How Players Raise Their Game When It Matters Most.” Medium, 3 June 2017, https://medium.com/@stevemagness/how-player-fatigue-will-impact-the-nba-finals-74d8003d84de.

\item “How Much Does Fatigue Matter in NBA Betting?.” OddsShark, https://www.oddsshark.com/\newline nba/how-much-does-fatigue-matter-nba-betting.

\item McMahan, I. “How sleep and jet-lag influences success in the travel-crazy NBA.” Guardian News \& Media Limited, 26 Oct. 2018, https://www.theguardian.com/sport/2018/oct/\newline 26/sleep-nba-effects-basketball.

\item “Professional teams are waking up to new methods of managing athlete fatigue.” Fatigue Science, 10 Oct. 2013, https://www.fatiguescience.com/blog/\newline  professional-teams-are-waking-up-to-new-methods-of-managing-athlete-fatigue.

\item Schuhmann, J. “Schedule Breakdown: How rest, road trips and other factors will impact 2021-22.” 2021 NBA Media Ventures, LLC, 22 Aug. 2021, https://www.nba.com/news/how-rest-road-trips-and-other-factors-played-out-in-2021-22-schedule.

\item Mullin, B. “Injury-Plagued NBA Draws Fewer TV Viewers.” 2021 Dow Jones \& Company, Inc., 27 Dec. 2019, https://www.wsj.com/articles/injury-plagued-nba-draws-fewer-tv-viewers-11577487612.

\item Pickman, B. “Report: NBA Could Issue Teams Fines of \$100K For Resting Players Throughout Season.” 2021 ABG-SI LLC. Sports Illustrated, 7 Dec. 2020, https://www.si.com/nba/2020/12/07/nba-resting-policy-load-management.

\item Edwards T., Spiteri T., Piggott B., Bonhotal J., Haff G.G., and Joyce C. “Monitoring and managing fatigue in Basketball.” Sports. 6: 19, 2018.

\item Seamon, B.A., and Harris-Love, M.O. “Clinical Assessment of Fatigability in Multiple Sclerosis: A Shift from Perception to Performance.” Front. Neurol. 7: 194, 2016. 

\item Stojanović, E., Stojiljković, N., Scanlan, A.T. et al. “The Activity Demands and Physiological Responses Encountered During Basketball Match-Play: A Systematic Review.” Sports Med 48: 111–135, 2018. 

\item Fox, J., Stanton, R., O'Grady, C., Teramoto, M., Sargent, C., and Scanlan, A. “Are acute player workloads associated with in-game performance in basketball?.” Biology of Sport, 2020.

\item García, A.C. “Training Load and Performance in Basketball: The More The Better?.” FC Barcelona Innovation Hub, 22 Oct. 2020, https://barcainnovationhub.com/training-load-and-performance-in-basketball-the-more-the-better/.

\item Holmes, B. “How fatigue shaped the season, and what it means for the playoffs.” ESPN Enterprises, Inc, 10 Apr. 2018, https://www.espn.com/nba/story/\_/
id/23094298/how-fatigue-shaped-nba
-season-means-playoffs.

\item Huyghe T., Scanlan A., Dalbo V., Calleja-González J. “The negative influence of air travel on health and performance in the National Basketball Association: A narrative review.” Sports. 6: 89, 2018.

\item Drew, M.K., Cook, J., and Finch, C.F. “Sports-related workload and injury risk:simply knowing the risks will not prevent injuries: Narrative review.” British Journal of Sports Medicine 50: 1306-1308, 2016.

\item Caparrós T., Casals M., Solana Á., and Pena J. “Low External Workloads Are Related to Higher Injury Risk in Professional Male Basketball Games.” J. Sports Sci. Med. 17: 289–297, 2018.

\item Lewis, M. “It's a hard-knock life: game load, fatigue, and injury risk in the National Basketball Association.” J. Athletic Train. 53, 503–509, 2018. 

\item Bresler, A. (2021). Nbastatr. R package version 0.1.1504. http://asbcllc.com/nbastatR/.

\item Fernandez, J. (2020). airball: Schedule \& Travel Related Metrics in Basketball. R package version 0.4.0. https://github.com/josedv82/airball.

\item Chodowski, P. (2021). NBAr. R package version 4.0.4. https://github.com/\newline PatrickChodowski/NBAr.

\item Abdelkrim, B.N., Castagna, C.; El Fazaa, S., and El Ati, J. 
“The Effect of Players' Standard and Tactical Strategy on Game Demands in Men's Basketball.” 
Journal of Strength and Conditioning Research 24 - Issue 10: 2652-2662, 2010.

\item Abdelkrim, B.N., El Fazaa, S., and El Ati J. “Time–motion analysis and physiological 
data of elite under-19-year-old basketball players during competition.” British Journal of 
Sports Medicine 41:69-75, 2007.

\item Lin, Mingfeng \& Lucas, Henry \& Shmueli, Galit. (2013). Too Big to Fail: Large Samples and the p-Value Problem. Information Systems Research. 24. 906-917. 10.1287/isre.2013.0480. 

\end{enumerate}

\end{flushleft}
\end{multicols*}

\end{document}